\documentclass[prl,twocolumn,amsmath,amssymb,bibnotes,aps,longbibliography]{revtex4-2}

\usepackage{graphicx}
\usepackage{dcolumn}
\usepackage{bm}

\usepackage[utf8]{inputenc}

\usepackage{times}
\usepackage{units}
\usepackage[mediumspace,mediumqspace,squaren]{SIunits}
\usepackage{physics}
\usepackage[english]{babel}
\usepackage{stmaryrd}
\usepackage{amssymb}
\usepackage{amsmath}
\usepackage{dsfont}
\usepackage{xcolor}
\usepackage[normalem]{ulem}

\newcommand{\Fig}{Fig.\@ }

\addto\captionsenglish{}

\usepackage{hyperref}
\hypersetup{
colorlinks=true, 
linkcolor=[RGB]{45,48,146}, 
citecolor=[RGB]{45,48,146}, 
filecolor=[RGB]{45,48,146}, 
urlcolor=[RGB]{45,48,146}}

\begin{document}
\title{Loss-Tolerant Quantum Correlations Between Atomic and Photonic Qubits}

\author{Leonardo~Ruscio$^{1}$}
\email[To whom correspondence should be addressed. Email: ]{leonardo.ruscio@mpq.mpg.de}
\author{Love~A.~Pettersson$^{3}$}
\author{Philip~Thomas$^{1,2}$}
\author{Stefano~Paesani$^{4}$}
\author{Olivier~Morin$^{1}$}
\author{Anders~S.~Sørensen$^{3}$}
\author{Gerhard~Rempe$^{1}$}

\affiliation{$^{1}$Max-Planck-Institut f{\"u}r Quantenoptik, Hans-Kopfermann-Strasse 1, 85748 Garching, Germany}
\affiliation{$^{2}$Fakultät für Physik, Ludwig-Maximilians-Universität München, 80799 München, Germany}
\affiliation{$^{3}$Center for Hybrid Quantum Networks (Hy-Q), Niels Bohr Institute, University of Copenhagen, Jagtvej 155A, 2200 Copenhagen N, Denmark.}
\affiliation{$^{4}$NNF Quantum Computing Programme, Niels Bohr Institute, University of Copenhagen, Blegdamsvej 17, 2100 Copenhagen, Denmark.}

\date{\today}

\begin{abstract}

\noindent Quantum physics forbids cloning of a qubit. Therefore, a qubit is irreversibly lost upon loss of the information carrier. To mitigate this physical loss, it has been proposed to distribute the logical information across an entangled set of physical qubits, some of which can be lost. Previous experiments illustrated this concept, but failed to demonstrate a gain in the information-retrieval efficiency. Here we address this challenge with photons for which the total loss, from generation to detection, is below the loss-tolerance threshold of $50\%$. Using two single-atom emitters in an optical cavity, we encode information into the photons of a graph state and observe an efficiency gain for retrieving quantum correlations between the atoms and the photons for single-qubit measurements in the Pauli $X$ and $Z$ bases. Although technical errors reduce the probability to obtain the \textit{correct} result, we achieve average fidelities as high as $\gtrsim\!70\%$ without and $\gtrsim\!75\%$ with error detection, demonstrating entanglement between the atomic and encoded photonic qubits. Lowering the encoding errors is a next natural step toward loss-tolerant photonic quantum-information processing.

\end{abstract}

\maketitle

\noindent Entanglement is arguably the most counterintuitive concept of quantum physics~\cite{Einstein1935}. It describes the observation that outcomes of measurements performed on a set of particles can be more strongly correlated than allowed by  classical physics~\cite{Bell1964}. Over the past few decades, experiments using pairs of entangled particles have not only explored fundamental quantum phenomena such as nonlocality~\cite{Brassard2023}, but have also established entanglement as a resource for new applications within the second quantum revolution~\cite{Dowling2003}.

Nevertheless, the generation, characterization, and control of entanglement in multi-qubit systems remains a central challenge~\cite{Horodecki2009}. In this context, graphs have emerged as a powerful language for visualizing a certain class of multipartite entangled states: \emph{graph states}. This formalism provides a pictorial representation of their entanglement structure. A key feature is their description in terms of a set of multi-qubit operators known as stabilizers~\cite{Hein2006}. The stabilizer framework can be efficiently employed to detect and correct qubit errors~\cite{Gottesman1997}, including physical loss of the qubit carrier~\cite{Ralph2005, Varnava2006, Buterakos2017, Borregaard2020, Love2025}. The ability to recover information despite loss is a particularly important feature, as qubits cannot be copied and held in reserve~\cite{Wootters1982}. Graph states are therefore a powerful resource for quantum information processing based on photons, for which loss is the dominant limitation. However, despite the tremendous progress achieved with photonic platforms~\cite{Walther2005, LuPNAS2008, Bell2014, Tomm2021, Martinez2022, Thomas2022, Cogan2023, Thomas2024, Meng2024,Ding2025,Huet2025,OSullivan2025}, an experimental demonstration that quantum information can survive photon loss has so far been missing.

\begin{figure}[b]
    \centering
    \includegraphics[]{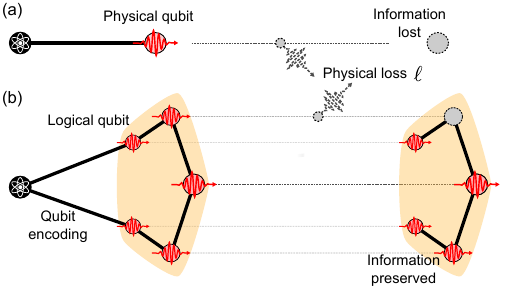}
    \caption{Core idea for loss tolerance. Black circles denote atomic qubits; light-red and gray circles denote available and lost photonic qubits, respectively. (a), A single photonic qubit entangled with one atom irreversibly loses the encoded information upon physical loss. (b), a logical qubit is encoded across five entangled photonic qubits, preserving the information despite the loss of one or more photons.} 
    \label{fig:idea}
\end{figure}

Here we report on an experiment in which we use a graph state to preserve quantum correlations between an atomic and a logical multi-photon qubit even when the information carriers, the photons, are subjected to loss. More specifically, instead of encoding information in a single photon (\Fig \ref{fig:idea}(a)), we distribute the qubit across multiple photons which serve as a resource that carries the logical information (\Fig \ref{fig:idea}(b)). We use a hexagonal graph state, perform logical single-qubit measurements in the $X$ and $Z$ Pauli bases, correlate the resulting logical photonic outcomes with measurements performed on the atomic qubit, and compare the success probability with that of single-photon encoding. Our experiment features a total physical loss in the generation, propagation, manipulation, and detection of a single photon below $50\%$ and thereby enters a regime where the logical loss is allowed to be below the physical loss~\cite{Varnava2006}. In fact, for both $X$ and $Z$ measurements, we obtain a higher detection probability of the quantum correlations using the multi-photon logical qubit than using a single-photon physical qubit. We note, however, that technical imperfections in the generation of the graph state introduce additional Pauli errors, with the consequence that the encoding does not increase the overall probability of obtaining the \textit{correct} logical measurement outcome. Nevertheless, we demonstrate genuine entanglement between the atomic and encoded photonic qubit with a fidelity $F\geq70.5\%$.

\begin{figure}
    \centering
    \includegraphics[]{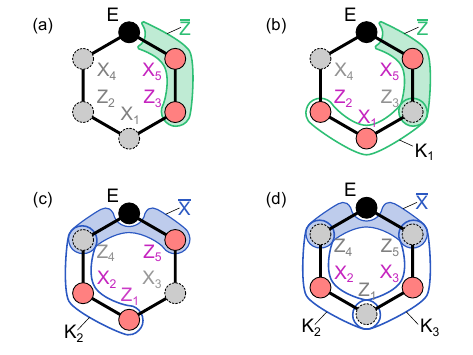}
    \caption{Reconstruction procedure for selected loss patterns. A measurement of $Z_E$ ($X_E$) on the encoding qubit $E$ is correlated with the logical operators $\overline{Z}=Z_3X_5$ ($\overline{X}=Z_4Z_5$) indicated by green (blue) shaded areas. Here, the photonic qubits are labeled from $1$ to $5$, with lost photons in gray. Encircling lines indicate stabilizer generators $\{K_i\}_{i=1,2,3}$. These have eigenvalue $K_i=1$ in the encoded state. (a) If photons $1$, $2$ and $4$ are lost, $\overline Z$ is obtained by measuring $Z_3$ and $X_5$. (b) If photons $3$ and $4$ are lost, we use $\overline{Z}K_1=X_1Z_2X_5$ to recover $\overline{Z}$ by measuring $X_1$, $Z_2$ and $X_5$. (c) Same loss pattern as in (b) but for recovery of $\overline{X}$ by using $\overline{X}K_2=Z_1X_2Z_5$ to reduce it to a measurement of $Z_1$, $X_2$ and $Z_5 $. (d) Recovery of $\overline{X}$ from measurements of only two qubit operators $X_2$ and $X_3$ by exploiting $\overline{X}K_2K_3=X_2X_3$.}
    \label{fig:protocol}
\end{figure}

\textit{Encoding of the photonic logical qubit} --- Graph states are a specific class of entangled multi-qubit states that can be visualized using graphs. Each vertex represents a qubit, and edges denote entangling controlled-$Z$ gates between the respective vertices. A graph state is an eigenstate of the stabilizers generated by the operators $K_{i}=X_{i}\prod_{k\in\mathcal{N}_{i}}Z_{k}$ with eigenvalue +1~\cite{Hein2004PRA, Hein2006}. Here, $X_i$ and $Z_k$ denote Pauli operators, the indices $i$ and $k$ label the vertices, and $\mathcal{N}_{i}$ represents the neighborhood of the $i$-th vertex. For instance, $K_1 = X_1 Z_2 Z_3$ for a qubit 1 connected to qubits 2 and 3 (\Fig \ref{fig:protocol}(b)).

Graph states can be used to implement measurement-based quantum error correction with graph codes~\cite{Hein2006}. In particular, as depicted in \Fig \ref{fig:idea}(b), a graph code encoding a logical qubit can be constructed by considering one of the qubits as an \emph{encoding qubit} (black circle) and the remaining qubits (light-red circles) as  carriers of the logical qubit (highlighted as shaded region). A logical state can be encoded by measuring the encoding qubit, effectively teleporting its state into the logical qubit~\cite{Bell2023}. The logical operators and stabilizers of the graph code are obtained from the graph structure (see Supplemental Material (SM) for details). Specifically, for the hexagonal graph code considered in this work, shown in \Fig \ref{fig:protocol}, we have the logical operators $\overline{X} = Z_4 Z_5$ and $\overline{Z} = X_5 Z_3$ (or $X_4 Z_2$), and stabilizer generators $\{ Z_5X_3Z_1, Z_3X_1Z_2, Z_1X_2Z_4, X_5Z_3Z_2X_4 \}$. The logical operators $\overline{X}$ and $\overline{Z}$ acting on the graph code are equivalent to the Pauli operators $X$ and $Z$ acting on a single qubit.

An important property of the hexagonal graph code is its ability to correct photon loss when performing the logical measurements. This is done by deforming its logical operators through multiplication with stabilizers, an operation which can provide other valid logical operators with no support on lost qubits~\cite{Gottesman1997}. One of the simplest cases occurs when photonic qubits $1$, $2$ and $4$ are lost, but not qubits $3$ and $5$ (\Fig \ref{fig:protocol}(a)). In this case, $\overline{Z} = X_5 Z_3$ tells us that a $Z$ measurement on qubit $3$ and an $X$ measurement on qubit $5$ can directly recover the encoded logical $\overline{Z}$ information. This is equal to the encoding $\overline{Z}=Z_E$, where subscript $E$ denotes the encoding qubit, since the original graph had $K_5=Z_EZ_3X_5=1$. By symmetry, the same information could also be extracted by measuring $Z_2X_4$, and this resilience is the reason for the loss tolerance. A less intuitive example is illustrated in \Fig \ref{fig:protocol}(b). If we attempt to measure $\overline{Z}$ and photonic qubit $3$ is lost, we can remove its support on this qubit by multiplying $\overline{Z}$ with $K_1$, obtaining an equivalent logical operator $\overline{Z} K_1 = X_5X_1Z_2$. If qubits 1, 2, and 5 are alive, measuring this Pauli operator allows us to recover the logical $\overline{Z}$ information even when qubits 3 and 4 are lost.

Analogous procedures can be used to recover logical information for various other loss configurations, such as those exemplified in \Fig \ref{fig:protocol}. For both complementary logical operators $\overline{Z}$ and $\overline{X}$, there are in total 2 configurations with 3 lost photons, 8 configurations with 2 lost photons, 5 configurations with 1 lost photon, and 1 configuration with no photon lost. This provides a probability to fail the recovery of the logical Pauli information given by

\begin{equation}
    \label{eq:log_loss}
    \overline{\ell} =1 - (1-\ell)^5 -5(1-\ell)^4\ell - 8 (1-\ell)^3 \ell^2 - 2(1-\ell)^2\ell^3,
\end{equation}
where $\ell$ represents the probability of losing each physical photon and $\overline{\ell}$ the logical loss probability, which turns out to be equal for an $\overline{X}$ and $\overline{Z}$ measurement.

Various graphs can be used to implement loss correction codes, the first one proposed being the tree state \cite{Varnava2006}. Ring graphs have recently been shown to be a powerful resource for long distance communication~\cite{Love2025} and are compatible with our experimental capabilities \cite{Thomas2024}. Equation (1) shows that for the hexagonal ring, the break-even point, i.e., the loss value below which the logical loss probability is lower than the physical loss, is as high as $50\%$. If losses are below $50\%$, the effect of photon loss is mitigated through encoding the information in the hexagonal graph code, indicating access to a loss-tolerant regime. Remarkably, this 50\% loss threshold is the maximum possible~\cite{Bell2023}: a loss correction code that would work for physical loss above 50\% would allow measuring $X$ and $Z$ simultaneously, violating fundamental principles of quantum mechanics, especially the no-cloning theorem.

\textit{Experimental code implementation} --- The core of the experiment and the hexagonal graph state generation protocol are illustrated in \Fig 2 of the SM, and described in greater detail in Ref.~\cite{Thomas2024}. We generate our photonic graph state using two $^{87}$Rb atoms trapped symmetrically about the center of a high-finesse optical cavity. Photon emission is performed by vacuum-Stimulated Raman Adiabatic Passage (vSTIRAP) on the $D_2$ line of Rubidium~\cite{MorinPRL2019}, and the photons are emitted into the cavity mode with their polarization being entangled with the atomic spin. The vSTIRAP process can be applied in two different configurations: globally, with a large beam interacting with both atoms and inducing photon emission simultaneously, and locally, addressing the two atoms independently and at different times with a tightly focused beam. We use the \emph{global} vSTIRAP to enable a cavity-assisted fusion mechanism~\cite{Thomas2024} that leaves the atoms in an entangled state (SM \Fig 2(b)), while the \emph{addressing} vSTIRAP generates individual photons entangled with the respective atom (SM \Fig 2(a)). Interleaving photon emissions with $\pi/2$ Raman rotations on the atomic qubit, we can produce linear cluster states~\cite{Thomas2022} from the two atoms independently. The photons are measured with a polarization-resolving detection setup  able to dynamically switch between $X$ and $Z$ measurement bases. 

To generate the hexagonal graph state we first entangle the atoms by performing cavity-assisted fusion using the global vSTIRAP. The two photons generated in the process are indistinguishable in the spatial and temporal degrees of freedom, which leads to erasure of which-path information. Therefore, an anti-correlated measurement outcome in the $Z$ basis heralds the entanglement of the atoms. We then use the addressing vSTIRAP to sequentially generate one photon from each atom. With a first $\pi/2$ rotation on the atoms, we generate a 3-qubit linear cluster state with two redundantly encoded photonic qubits. One photonic qubit is removed from the graph via an $X$ measurement. Repeating independent photon emissions and global $\pi/2$ rotations two more times, we obtain a 7-qubit linear cluster state with the two atoms at each end. A second and final fusion generates the 6-qubit hexagonal graph state, where the two atoms redundantly encode a single computational qubit (in a two-qubit repetition code), which we use as the encoding qubit for the graph code. An important feature of our graph code generation is that, in contrast to previous implementations of quantum error correction of photon loss~\cite{LuPNAS2008,Bell2014}, it does not rely on postselection. That is, all processes used in its generation are either deterministic (spin-photon entanglement generation) or probabilistic but heralded (fusion operations). Conditioned on the successful heralding of the two fusions and the first $X$-measured photon, all output photon combinations are accepted with no postselection, enabling us to rigorously test the loss-tolerant properties of the generated graph code.

In the experiment we achieve a source-to-detection efficiency of $52.6\%$, which surpasses the $50\%$ break-even point for the hexagonal code and enables us to enter a loss-tolerant regime. The photon emission efficiency at the output of the cavity is $69\%$ (see cavity parameters in the SM and~\cite{MorinPRL2019}). The detectors (superconducting nanowire single photon detectors) have an estimated efficiency of $95\%$. From the cavity to the detectors the efficiency of the optical path is around $86\%$ (imperfect anti-reflection coatings, finite fiber mode-matching, etc.). For a more detailed breakdown of the efficiency budget, see the SM.

\begin{figure}
    \centering
    \includegraphics[]{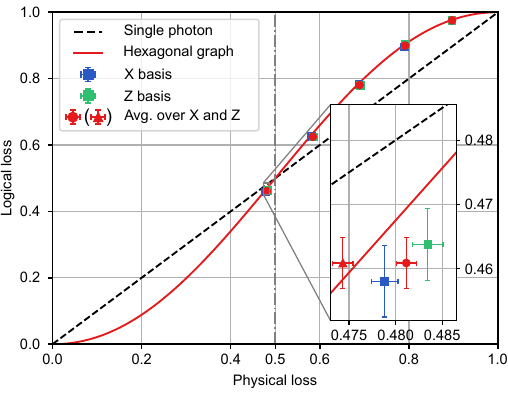}
    \caption{Loss tolerance. The black dashed line and the red curve represent the theoretical logical loss as a function of the physical loss of a qubit encoded in a single photon and a hexagonal graph state, respectively. The inset shows how the experimental data for the hexagonal state surpass the logical loss of a single photon. The green and blue points are logical $Z$ and $X$ measurements while the red circle is the average of the two. The experimental points laying below the theory curve can be explained by the difference in efficiency between the photons composing the graph (see SM for details). For the red triangle we define the abscissa by the physical loss of the most efficient photon. All error bars are one-sigma confidence intervals.}
    \label{fig:Logical loss theory and results}
\end{figure}


\textit{Loss tolerance analysis} --- For each hexagonal graph state generation attempt, conditioned on the success of the two fusion gates, we use the stabilizers and logical operators of the graph code to access the measurement outcome of the encoding qubit. As discussed before, this is feasible for specific subsets of detected photons depending on the logical measurement basis. Combinations for which the information of the encoding qubit cannot be recovered constitute a failure of the protocol. The average number of such failures, as a function of the physical loss, defines the logical loss. The theoretical logical loss function is given in Eq.~\eqref{eq:log_loss} and shown in \Fig \ref{fig:Logical loss theory and results}, alongside the single-qubit line. The two curves intersect at a physical loss value of $0.5$. Demonstrating that, for a physical loss lower than $0.5$, the logical loss falls below that of a single photon, represents the central result of this work. 

A zoomed-in view of this result is shown in the inset of \Fig \ref{fig:Logical loss theory and results}. We independently measure the logical loss in the $X$ (blue square) and $Z$ (green square) bases, and the final value is calculated as the average of the $X$ and $Z$ results (red circle). The physical loss is extracted from the same datasets and defined as the ratio of undetected photons to the number of generation attempts, regardless of the photon's position in the graph state. For the $X$ basis we measure a physical loss of $0.479(1)$ and a logical loss of $0.458(6)$, for $Z$ we measure a physical loss of $0.483(2)$ and a logical loss of $0.464(6)$. Finally, we calculate the average physical and logical loss obtaining $0.481(1)$ and $0.461(4)$, respectively. To unambiguously demonstrate loss tolerance, given that the physical loss of individual photons can change due to imperfections in the atomic control processes, we include another data point (red triangle). In this case, the logical loss is still calculated as the average of the $X$ and $Z$ data, but the physical loss is taken to be the minimum among all photons ($0.474(1)$). In both cases, we demonstrate that the logical loss of the hexagonal graph state is significantly lower than the single-photon loss, with statistical significance exceeding three standard deviations.

Additionally, to show agreement between the predicted logical loss and the measured data over a wider range of loss values, we artificially vary the physical loss by discarding, with a certain probability, the photon detection events during data processing. The resulting data points closely follow the theoretical model as shown in \Fig \ref{fig:Logical loss theory and results}.


\begin{figure}
    \centering
    \includegraphics[]{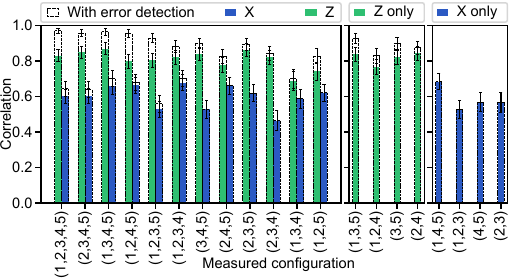}
    \caption{Correlations for recovered logical operators. Blue (green) bars represent the correlation between the encoded and recovered logical operators in the $X$ ($Z$) basis, for the different successful measurement configurations. Dashed bars represent the corresponding value when error detection is used to improve the error rate by filtering out the cases where errors are flagged by the measured syndrome. The left plot includes configurations for which the retrieval can be successful in both the $Z$ and $X$ bases, while the central and right plots show correlations for configurations that succeed only in the $Z$ basis and only in the $X$ basis, respectively.} 
    \label{fig:Fidelity}
\end{figure}

\textit{Fidelity and entanglement discussion} --- While single-qubit measurements of the logical photonic qubit provide logical information with an efficiency higher than for single-photon encoding, noise affecting the atomic and photonic qubits can still induce logical errors in the recovered information, lowering the entanglement fidelity between the atoms and the photons with respect to the single-photon case. The dominant noise sources are: magnetic field fluctuations causing decoherence in the atomic spin states during the generation of the graph state; distinguishability between the photons used in the fusion operation; imperfect Raman rotations on the atoms; and imperfect polarization manipulations of the photons. Since the resulting reduction in fidelity exceeds the gain in efficiency for our experimental parameters, the encoded qubit has a lower rate of \textit{correct} outcomes than a single photon. An important future milestone is therefore the demonstration of an increase in efficiency that outweighs any reduction in fidelity, i.e. an increase in the overall probability of obtaining the \textit{correct} measurement outcomes.

In \Fig \ref{fig:Fidelity}, we report the correlators between the encoding qubit operators $Z_E$ and $X_E$ acting on the atoms and the recovered logical operators acting on the photons $\overline{Z}$ and $\overline{X}$. That is, $\langle Z_E\overline{Z}\rangle$ and  $\langle X_E\overline{X}\rangle$ for the $Z$ and $X$ bases, respectively. Considering all successful configurations (including those shown in the center and right plot of \Fig \ref{fig:Fidelity}, where only $Z$ or only $X$ can be recovered), we obtain a value of $\langle Z_E \overline{Z}\rangle = 0.81(1)$ in the $Z$ basis, and $\langle X_E \overline{X}\rangle = 0.60(1)$ in the $X$ basis. Assuming fair sampling on configurations where only $\overline{Z}$ or only $\overline{X}$ is measurable (see SM), these correlations can be thought of as the creation of a Bell state between the input qubit and the logical photonic qubit with a fidelity~\cite{GuhneToth2005} of $F \geq (\langle Z_E \overline{Z}\rangle + \langle X_E \overline{X}\rangle)/2 = 0.705(8)$.

The values above are obtained by considering, for each successful event, the product of the measurements corresponding to the logical operator with the maximum fidelity compatible with the measurement configuration of that event, and averaging over all successful events. However, if enough photons survive, the logical operator can be reconstructed from multiple products of the Pauli measurements of the recovered photons, all equivalent up to stabilizers of the graph code. In these cases, error detection can be used to reduce the probability of logical errors. In particular, if not all equivalent logical operators provide the same value, a syndrome flagging the presence of at least one error on the code qubits is obtained. In error detection, we filter the flagged cases out when they appear. The results for each measurement configuration and basis are shown as dashed bars in \Fig \ref{fig:Fidelity}. Error detection improves correlations in the $Z$ basis significantly to $\langle Z_E \overline{Z}\rangle_\text{ED} = 0.889(8)$, while in the $X$ basis it provides a smaller average improvement to $\langle X_E \overline{X}\rangle_\text{ED} = 0.61(1)$. If we again assume fair sampling for the loss configurations where we can only measure $Z$ or $X$, we get a fidelity to a Bell state of $F_\text{ED} \geq 0.750(7)$. One can see from \Fig \ref{fig:Fidelity} that, as expected, the improvement from error detection is generally higher for the configurations with more photons being detected, because these configurations provide more code stabilizers that can be used for syndrome detection. However, due to the filtering out of noisy events flagged by the syndromes, the probability of losing the logical operator is increased to $0.521(6)$ and $0.478(6)$ for the $Z$ and $X$ bases, respectively.

Furthermore, one might want to restrict the analysis to measurement configurations that allow recovery of the logical information in both $Z$ and $X$ bases (configurations in the left plot of \Fig \ref{fig:Fidelity}), as could be required in quantum secure applications. Indeed, if we restrict ourselves to these configurations, the equivalence of the encoded state to a two-level qubit requires no fair sampling assumption. Considering only these configurations, we obtain the same correlations as in the first case, providing a direct verification of entanglement between the atoms and the encoded photonic qubit without the fair sampling assumption. The loss probabilities in this scenario are $0.593(6)$ and $0.580(6)$ for the $Z$ and $X$ basis, respectively, increased from previous cases due to neglecting the configurations where only $X$ or only $Z$ can be recovered. We emphasize, however, that for, e.g., loophole-free violation of Bell's inequality and fault-tolerant quantum computation, the loss configurations where only $Z$ or $X$ can be extracted still provide useful information (see SM), justifying the smaller logical loss rate for some applications.

\textit{Summary and outlook} --- We have implemented an error correcting code capable of retrieving quantum correlations despite loss. For a physical loss of $48.1(1)\%$, we have measured a smaller logical loss of $46.1(4)\%$. This tolerance against loss has been observed with graph-state encoding and for single-qubit measurements in the Pauli $X$ and $Z$ bases. Our scheme represents an important primitive for loss-tolerant quantum information processing, but the demonstration of a full-fledged architecture may require additional loss-tolerant operations such as single-qubit measurements in arbitrary bases and re-encoding of the logical qubit into a memory qubit. Future experiments could push the source-to-detection efficiency even further, e.g., by increasing the photon-emission efficiency, to demonstrate loss tolerance for quantum secure applications. Furthermore, more complex graph states like cubes or hypercubes~\cite{Bell2023} are worth investigating for increased loss tolerance, and more generally for quantum computing~\cite{Bartolucci2023,Loebl2023}. However, this will require more photon emitters~\cite{Li2022}. This could be achieved either with multiple cavities connected by traveling photons~\cite{Daiss2021} or multiple atoms in the same cavity~\cite{Hartung2024}. Alternatively, optical delay lines could be used to fuse photons emitted by a single atom at different times \cite{Loebl2025}.

\textit{Acknowledgments} --- This work was supported by Bundesministerium f\"{u}r Bildung und Forschung via the Verbund QR.X (16KISQ019), Bundesministerium f\"{u}r Forschung, Transfer und Raumfahrt via the Verbund QR.N (16KIS2189), Deutsche Forschungsgemeinschaft under Germany’s Excellence Strategy – EXC-2111 – 390814868, European Union’s Horizon 2020 research and innovation programme via the project Quantum Internet Alliance (QIA, GA No. 820445), Danmarks Grundforskningsfond (DNRF 139, Hy-Q Center for Hybrid Quantum Networks), and Danmarks Innovationsfond (IFD1003402609,
FTQP). S.P. acknowledges funding from the VILLUM FONDEN research grant No.~VIL60743, from the ERC Starting Grant ``ASPEQT'' (No. 101221875), and support from the NNF Quantum Computing Programme. L.A.P. acknowledges support from Novo Nordisk Foundation (Challenge project “Solid-Q”).

\let\oldaddcontentsline\addcontentsline
\renewcommand{\addcontentsline}[3]{}

\end{document}


\title{Supplemental Material: Loss-Tolerant Quantum Correlations Between Atomic and Photonic Qubits}

\author{Leonardo~Ruscio$^{1}$}
\email[To whom correspondence should be addressed. Email: ]{leonardo.ruscio@mpq.mpg.de}
\author{Love~A.~Pettersson$^{3}$}
\author{Philip~Thomas$^{1,2}$}
\author{Stefano~Paesani$^{4}$}
\author{Olivier~Morin$^{1}$}
\author{Anders~S.~Sørensen$^{3}$}
\author{Gerhard~Rempe$^{1}$}

\affiliation{$^{1}$Max-Planck-Institut f{\"u}r Quantenoptik, Hans-Kopfermann-Strasse 1, 85748 Garching, Germany}
\affiliation{$^{2}$Fakultät für Physik, Ludwig-Maximilians-Universität München, 80799 München, Germany}
\affiliation{$^{3}$Center for Hybrid Quantum Networks (Hy-Q), Niels Bohr Institute, University of Copenhagen, Jagtvej 155A, 2200 Copenhagen N, Denmark.}
\affiliation{$^{4}$NNF Quantum Computing Programme, Niels Bohr Institute, University of Copenhagen, Blegdamsvej 17, 2100 Copenhagen, Denmark.}

\date{\today}

\maketitle

\tableofcontents

\subsection{Conventions}

In the following we use the conventions:

\begin{itemize}
    \item $\ket{F=1,m_F=1} = \ket{0}_A$
    \item $\ket{F=1,m_F=-1} = \ket{1}_A$
    \item $\ket{R} = \ket{0}_P$
    \item $\ket{L} = \ket{1}_P$
    \item $\ket{+} = \frac{1}{\sqrt{2}}(\ket{0}+\ket{1})$
    \item $\ket{-} = \frac{1}{\sqrt{2}}(\ket{0}-\ket{1})$
    \item $\ket{0}_A \xrightarrow{\pi/2} \ket{+}_A$
    \item $\ket{1}_A \xrightarrow{\pi/2} \ket{-}_A$
\end{itemize}
where $F$ denotes the total-angular-momentum quantum number of the Rubidium atom; $m_F$ the Zeeman quantum number $m_F = -F, \dots , F$; $R$ and $L$ the right and left circular polarizations of a photon; and the indices $A$ and $P$ an atomic and a photonic state, respectively.

\begin{figure}
    \centering
    \includegraphics[]{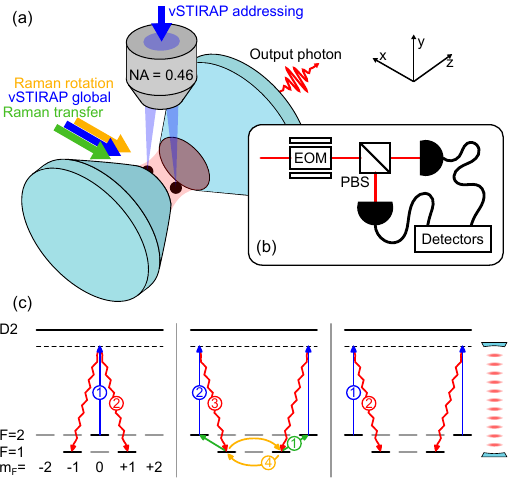}
    \caption{Experimental setup: (a), Schematic (not to scale) of the cavity QED system. Two atoms are trapped at the center of an optical cavity (trapping beams not shown). They can be addressed either individually, using a laser beam tightly focused by a high numerical aperture objective ($NA=0.46$), or globally, with laser beams (Raman, vSTIRAP) propagating perpendicularly to the cavity axis. Photons emitted into the cavity mode are directed to the detection setup, sketched in (b). A polarizing beam splitter (PBS) discriminates between orthogonal polarizations, while an electro-optic modulator (EOM) enables switching between two measurement bases. The photons are detected using two superconducting nanowire single-photon detectors (SNSPDs). (c), Level scheme and pulse sequence. The left and right schemes show the initial and final cavity assisted fusions used in the graph state generation protocol, where the blue arrows represent the global vSTIRAP control pulse applied to the two atoms simultaneously. The central scheme shows the repeated photon emission sequence: 1. Raman pulse to transfer the population from the $F=1$ to the $F=2$ hyperfine levels. 2. Addressing vSTIRAP control pulse on the two atoms independently. 3. Photon emission in the cavity mode. 4. Global $\pi/2$ Raman rotation to generate the linear cluster state as described in the main text.}
    \label{fig:full_setup}
\end{figure}

\subsection*{Experimental setup}

Our experimental setup has been described in detail in~\cite{Thomas2022, Thomas2024}, and is shown in \Fig \ref{fig:full_setup}. Here we only provide a summary of the most relevant features. The heart of the experiment is a high-finesse optical cavity (with parameters displayed in \Tab \ref{tab:cavity_parameters}), in which we trap two $^{87}$Rb atoms in a two-dimensional standing-wave dipole potential formed by a $1064$nm retro-reflected laser beam perpendicular ($x$-axis) to the cavity axis ($z$-axis) and a standing-wave intracavity laser at $772$nm. To manipulate the internal electronic state of the atoms, we use several laser beams propagating perpendicularly with respect to the cavity axis. We use them to initialize all the atoms in a specific state with optical pumping, to generate photons by a vacuum stimulated Raman adiabatic passage technique (vSTIRAP)~\cite{MorinPRL2019}, and to coherently drive Raman transitions for the implementation of qubit rotations or qubit transfers. Photon generation is performed on the D2 line with the control pulse (blue arrows in \Fig \ref{fig:full_setup} c) addressing the $\ket{F=2, m_F = \pm2} \rightarrow \ket{F'=2,m'_F=\pm2}$ transition. The photons are emitted into the cavity mode (red arrows) which is detuned by $\sim 150$MHz from the $\ket{F=1,m_F=\pm1} \rightarrow \ket{F'=1,m'_F=\pm1}$ transition. The cavity is unbalanced, and the emitted photons leave the cavity predominantly through one side. They are measured with a polarization resolving detection setup which is coupled via free-space optics to the output of the cavity.

\begin{table}
    \centering
    \begin{tabular}{|l|c|r|c|}
    \hline
    \textbf{Description} & \textbf{Symbol} & \textbf{Value} & \textbf{Unit} \\
    \hline
    Cavity length & $d$ & $500$ & µm \\
    Cavity mode waist & $w_0$ & $30$ & µm \\
    Mirror 1 transmittivity & $T_1$ & $100$ & ppm \\
    Mirror 2 transmittivity & $T_2$ & $4$ & ppm \\
    Finesse & $\mathcal{F}$ & $60000$ & - \\
    Atom-cavity coupling & $g/2\pi$ & $5.4$ & MHz \\
    Cavity decay rate & $\kappa /2\pi$ & $2.7$ & MHz \\
    Atom decay rate to free space & $\gamma /2\pi$ & $3.0$ & MHz \\
    Cooperativity & $C$ & $1.8$ & - \\
    \hline
    \end{tabular}
    \caption{Table containing the most relevant cavity parameters. The atom-cavity coupling parameter $g$, and thus the cooperativity $C=g^2/2\kappa\gamma$, depend on the Clebsch-Gordon coefficient of the photon emission transition. In our case, $\ket{F=1,m_F=\pm1}\longrightarrow\ket{F'=2,m'_F=\pm2}$ with a coefficient of $\sqrt{1/4}$.}
    \label{tab:cavity_parameters}
\end{table}

\begin{figure}
    \centering
    \includegraphics[]{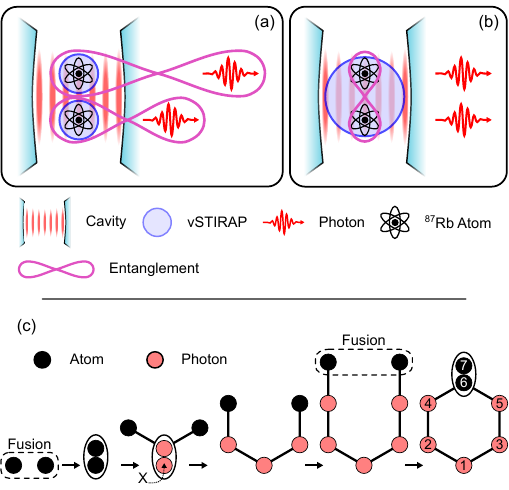}
    \caption{Experimental graph state generation. (a) and (b) show the two-photon emission mechanisms: in (a) the two atoms are addressed independently at different times and the emitted photons are entangled with the respective atoms. In the fusion mechanism (b), the vSTIRAP control pulse is global, and the two generated photons are indistinguishable. The result, depending on the polarization of the photons, is entanglement between the two atoms. (c), Hexagonal state generation scheme. After entangling the two atoms, a linear cluster state is grown from a succession of photon emissions and atomic state rotations. The first photon is measured in $X$ and, serving as a herald, can be removed from the graph. A final fusion step merges the ends of the linear cluster states to form a 6-qubit hexagonal graph state (made of 7 physical qubits).} 
    \label{fig:setup}
\end{figure}

\subsubsection*{Global photon emission and fusion mechanism}

To entangle the two atoms in the cavity (\Fig \ref{fig:setup} (b)), we use a global vSTIRAP control pulse propagating in the direction of the $1064$nm trap, and thus illuminating uniformly the trapped atoms. The photons are emitted in the same spatiotemporal mode. Upon detection of two photons with anti-correlated polarizations at opposite outputs of the PBS, due to the impossibility of distinguishing which atom emitted which photon, the atomic state is projected into the $\ket{\Psi^+} = \frac{1}{\sqrt{2}}(\ket{0,1}_A+\ket{1,0}_A)$ Bell state. The fidelity of the fusion depends on the indistinguishability of the photons, which is degraded due to off-resonant scattering and the finite atom-photon cooperativity. However, the fidelity can be improved by filtering on the difference in arrival time of the photons. We use here a maximal time difference of $250$ns for photons emitted from $\ket{F=2, m_F=0}$ and $400$ns for photons emitted from $\ket{F=2, m_F\pm2}$ (same as in~\cite{Thomas2024}).

\subsubsection*{Local photon emission and atom read-out}

We use a local vSTIRAP beam to independently address the two atoms trapped in the cavity (\Fig \ref{fig:setup} (a)). We can steer this addressing beam using an acousto-optical deflector. The beam has an elliptical profile, and it is tightly focused onto the atoms by means of a high numerical aperture objective. The elliptical shape, with the major axis aligned to the cavity axis, reduces crosstalk between adjacent atoms while providing a more uniform vSTIRAP intensity profile along the cavity axis. Along the $x$-axis we measure an addressing beam profile with a FWHM of $2\mu\mathrm{m}$. We use the addressing vSTIRAP beam to independently generate photonic graph states from the two atoms. In this way, we have access to two-dimensional topologies that cannot be obtained using only the global beam. We also use the addressing vSTIRAP beam to read out the atomic state at the end of the protocol. After the final fusion, we transfer the qubit to $\ket{F=2, m_F\pm2}$, and we generate two separated photons from the two atoms that are measured with the polarization resolving detection setup. The detection basis in this case is changed by applying a Raman rotation directly to the atoms. This protocol can be repeated until success (three times in our experiment), thereby increasing the read-out efficiency.

\begin{table}
    \centering
    \begin{tabular}{|l|r|r|r|}
    \hline
    Optical element & Nr. of elts. & Efficiency & Tot. efficiency \\
    \hline
    Narrow-band coated lenses & 4 & $0.998$ & $0.992$ \\
    Broad-band ach. doublets & 2 & $0.990$ & $0.980$ \\
    Mirrors & 10 & $0.999$ & $0.990$ \\
    Filters & 2 & $0.997$ & $0.994$ \\
    Waveplates & 6 & $0.997$ & $0.982$ \\
    EOMs & 2 & $0.990$ & $0.980$ \\
    PBS & 1 & $0.990$ & $0.990$ \\
    Mode matching & 1 & $0.980$ & $0.980$ \\
    Fibre tip & 1 & $0.998$ & $0.998$ \\
    Fibre length ($30$m) & 1 & $0.980$ & $0.980$ \\
    Fibre-fibre connection & 1 & $0.990$ & $0.990$ \\
    Detector efficiency & 1 & $0.950$ & $0.950$ \\
    \hline
    Tot. efficiency &  &  & 0.822 \\
    \hline
    \end{tabular}
    \caption{Table containing the expected efficiency values for each optical element used in the detection setup. Most of the provided numbers are based on the supplier's specifications, as measuring the efficiency of optical components at the percent level is not a trivial task. The setup was built using a low-power ($\sim 100$nW) probe light through the cavity to prevent damage due to the high finesse. At these powers one can only optimize the transmission but never accurately measure it.}
    \label{tab:efficiency_budget}
\end{table}

\subsubsection*{Optical source-to-detection efficiency}

\begin{figure}
    \centering
    \includegraphics[]{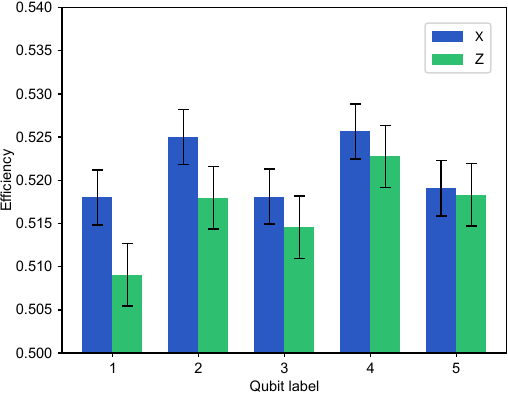}
    \caption{Source-to-detection efficiencies of the photonic qubits forming the hexagonal state (labeled 1-5 as in the main text). Each photon efficiency is measured in the $X$ (blue bars) and $Z$ (green bars) bases, giving slightly different results due to the fact that the two bases are measured in independent experimental runs. The error bars represent a $1\sigma$ error.}
    \label{fig:eff}
\end{figure}

The detection setup is carefully designed to maximize the optical efficiency. A detailed description of all optical components, along with their expected transmission efficiencies, is provided in table \ref{tab:efficiency_budget}. The resulting optical efficiency of the detection setup is $82\%$. It accounts only for optical losses and does not include loss mechanisms arising from the manipulation of internal atomic states. If we include the $69\%$ photon emission efficiency at the output of the atom-cavity system, obtained from the cooperativity $C=1.8$, we estimate a source-to-detection efficiency of $56\%$. It is important to remember that to generate the photons, we use two different vSTIRAP beams (see Experimental setup section above), a global and an addressing beam. While the frequency of both beams is accurately stabilized by a frequency-comb locking system, intensity fluctuations and polarization misalignment can adversely affect photon emission efficiency, especially in the case of the addressing beam. With the global vSTIRAP beam, we measure a maximum efficiency of $54.6\%$, which is close to the expected value of source-to-detection efficiency. With the addressing vSTIRAP beam, we measure a maximum efficiency of $53.2\%$. These values are measured independently of the experimental data used for the hexagonal graph code. If we consider the graph code data, the five photons composing the graph are generated with the addressing vSTIRAP beam. Each of them, due to imperfect atomic manipulations, has a different efficiency shown in \Fig \ref{fig:eff}. The maximum measured single photon efficiency in this case is $52.6\%$. The discrepancy between this value and the $53.2\%$ obtained independently is largely explained by the fact that, in the latter case, the photon is emitted only after careful preparation of the atomic population. In contrast, during the graph state generation sequence, the photon is emitted after a more complex series of Raman manipulations, which can drive the atomic population into undesired states. There are two main reasons why the efficiency of the addressing beam is lower than that of the global beam. First, the addressing beam is more tightly focused onto the atoms, making the photon emission process more sensitive to fluctuations in the position of the atom relative to the beam center. These fluctuations cause variations in the light shift induced on the atomic levels, resulting in a lower average coupling efficiency to the cavity mode. The second reason is that the power stabilization of the addressing beam is more challenging than for the global beam. This is because the pointing direction of the addressing dynamically follows the atoms, whereas the global remains fixed. Instabilities in the power stabilization scheme once again induce fluctuations in the light shift.

\subsubsection*{Atoms positioning}

In our experiment, we have control over the position of the atoms along the global beams propagation axis ($x$-axis) and along the cavity axis ($z$-axis). Along the $x$-axis, we can accurately control the position of the atoms' center of mass by shifting the phase of the $1064$nm standing-wave potential. In contrast, along the $z$-axis we can only probabilistically reposition the atoms by modulating the power of the $772$nm trap. In this way, there is a certain probability that, due to the red-detuned trapping potential, the atoms will be located in a more centred position. The position selection criteria are important to reach the high efficiencies required in the experiment. The distance between the atoms must be large enough to avoid cross-talk during the addressing vSTIRAP process, yet close enough to ensure maximal coupling to the cavity mode (waist $w_0=30\mu\mathrm{m}$). Along the $x$-axis we select atoms with a distance between $2$ and $12 \mu\mathrm{m}$. Along the $z$-axis, even though the addressing beam is elongated to ensure a uniform intensity profile, we select pairs of atoms with a maximum distance from the cavity center of $2\mu\mathrm{m}$.

\subsection*{Detailed experimental protocol}

In the following, we neglect all normalization factors. Referring to the level structure schematized in \Fig \ref{fig:full_setup} c, we start the protocol by preparing both atoms in the state $\ket{F=2,m_F=0}$ via optical pumping. We then entangle the atoms with the fusion mechanism described above. Upon successful measurement of two anti-correlated photons, we can write the atomic state as:

\begin{equation}
    \ket{\Psi}_A = \ket{0}_{A_1}\ket{1}_{A_2}+\ket{1}_{A_1}\ket{0}_{A_2}.
\end{equation}

\noindent We only continue the protocol if this BSM is successful; otherwise, the sequence is re-initialized. After the fusion of the two atoms, we produce the first pair of photons with the addressing vSTIRAP beam. We can write the atom-photon state as:

\begin{equation}
    \ket{\Psi}_{AP} = \ket{0}_{A_1}\ket{01}_P\ket{1}_{A_2}+\ket{1}_{A_1}\ket{10}_P\ket{0}_{A_2}.
\end{equation}

\noindent We now apply a $\pi/2$ rotation globally on the atomic qubits and we obtain:

\begin{equation}
    \ket{\Psi}_{AP} = \ket{+}_{A_1}\ket{01}_P\ket{-}_{A_2}+\ket{-}_{A_1}\ket{10}_P\ket{+}_{A_2}.
\end{equation}

\noindent We can consider the two photonic qubits as a single logical qubit ($\ket{01}=\ket{\overline{1}}$ and $\ket{10} = \ket{\overline{0}}$). We can then rewrite the state as:

\begin{equation}
    \ket{\Psi}_{AP} = \ket{+}_{A_1}\ket{\overline{1}}_P\ket{-}_{A_2}+\ket{-}_{A_1}\ket{\overline{0}}_P\ket{+}_{A_2},
\end{equation}

\noindent which, up to a local unitary operation, is equal to the state:
\begin{widetext}
\begin{equation}
    \ket{+\overline{0}+}+\ket{-\overline{1}-} = \ket{0\overline{0}0}+\ket{0\overline{0}1}+\ket{0\overline{1}0}-\ket{0\overline{1}1}+\ket{1\overline{0}0}+\ket{1\overline{0}1}-\ket{1\overline{1}0}+\ket{1\overline{1}1},
\end{equation}
\end{widetext}

\noindent representing a $3$-qubit linear cluster state with two photonic qubits encoding a single logical qubit at the centre of the cluster, and two atoms at the ends. The sequence continues alternating two more photon emissions (PP) and $\pi/2$ rotations. Removing one photon with an $X$ measurement, and dropping the $P$ of the photonic qubits, we can write:
\begin{widetext}
\begin{align*}
    \xrightarrow{eq.4} \;\;\; & \ket{+0+}+\ket{-1-}\\
    = \;\;\; & (\ket{0}+\ket{1})_{A_1}\ket{0}(\ket{0}+\ket{1})_{A_2}+(\ket{0}-\ket{1})_{A_1}\ket{1}(\ket{0}-\ket{1})_{A_2}\\
    = \;\;\; & \ket{0}_{A_1}\ket{+}\ket{0}_{A_2}+\ket{0}_{A_1}\ket{-}\ket{1}_{A_2}+\ket{1}_{A_1}\ket{-}\ket{0}_{A_2}+\ket{1}_{A_1}\ket{+}\ket{1}_{A_2}\\
    \xrightarrow{PP} \;\;\; & \ket{0}_{A_1}\ket{0+0}\ket{0}_{A_2}+\ket{0}_{A_1}\ket{0-1}\ket{1}_{A_2}+\ket{1}_{A_1}\ket{1-0}\ket{0}_{A_2}+\ket{1}_{A_1}\ket{1+1}\ket{1}_{A_2}\\
    \xrightarrow{\pi/2+Z_{A_1}} \;\;\; & \ket{-}_{A_1}\ket{0+0}\ket{+}_{A_2}+\ket{-}_{A_1}\ket{0-1}\ket{-}_{A_2}\\
    & +\ket{+}_{A_1}\ket{1-0}\ket{+}_{A_2}+\ket{+}_{A_1}\ket{1+1}\ket{-}_{A_2}\\
    = \;\;\; & (\ket{00}+\ket{01}-\ket{10}-\ket{11})_A\ket{0+0}\\
    + & (\ket{00}-\ket{01}-\ket{10}+\ket{11})_A\ket{0-1}\\
    + & (\ket{00}+\ket{01}+\ket{10}+\ket{11})_A\ket{1-0}\\
    + & (\ket{00}-\ket{01}+\ket{10}-\ket{11})_A\ket{1+1}\\
    \xrightarrow{PP} \;\;\; & \ket{0}_{A_1}(\ket{00+00}+\ket{00-10}+\ket{01-00}+\ket{01+10})\ket{0}_{A_2}\\
    + & \ket{0}_{A_1}(\ket{00+01}-\ket{00-11}+\ket{01-01}-\ket{01+11})\ket{1}_{A_2}\\
    - & \ket{1}_{A_1}(\ket{10+00}+\ket{10-10}-\ket{11-00}-\ket{11+10})\ket{0}_{A_2}\\
    - & \ket{1}_{A_1}(\ket{10+01}-\ket{10-11}-\ket{11-01}+\ket{11+11})\ket{1}_{A_2}\\
    \xrightarrow{\pi/2} \;\;\; & \ket{+}_{A_1}(\ket{00+00}+\ket{00-10}+\ket{01-00}+\ket{01+10})\ket{+}_{A_2}\\
    + & \ket{+}_{A_1}(\ket{00+01}-\ket{00-11}+\ket{01-01}-\ket{01+11})\ket{-}_{A_2}\\
    - & \ket{-}_{A_1}(\ket{10+00}+\ket{10-10}-\ket{11-00}-\ket{11+10})\ket{+}_{A_2}\\
    - & \ket{-}_{A_1}(\ket{10+01}-\ket{10-11}-\ket{11-01}+\ket{11+11})\ket{-}_{A_2}.
\end{align*}
\end{widetext}

\noindent As a final step, we perform again the fusion of the two atoms. Expanding the atomic state in the computational basis, selecting only the anti-correlated atomic states and omitting the atom subscript $A$, we obtain:

\begin{align*}
    \xrightarrow{\text{Fusion}} \;\;\; & \ket{00+001}+\ket{00-101}+\ket{01-001}+\ket{01+101}\\
    + & \ket{00+000}+\ket{00-100}+\ket{01-000}+\ket{01+100}\\
    - & \ket{00+011}+\ket{00-111}-\ket{01-011}+\ket{01+111}\\
    + & \ket{00+010}-\ket{00-110}+\ket{01-010}-\ket{01+110}\\
    - & \ket{10+001}-\ket{10-101}+\ket{11-001}+\ket{11+101}\\
    + & \ket{10+000}+\ket{10-100}-\ket{11-000}-\ket{11+100}\\
    + & \ket{10+011}-\ket{10-111}-\ket{11-011}+\ket{11+111}\\
    + & \ket{10+010}-\ket{10-110}-\ket{11-010}+\ket{11+110}
\end{align*}

\begin{figure}
    \centering
    \includegraphics[]{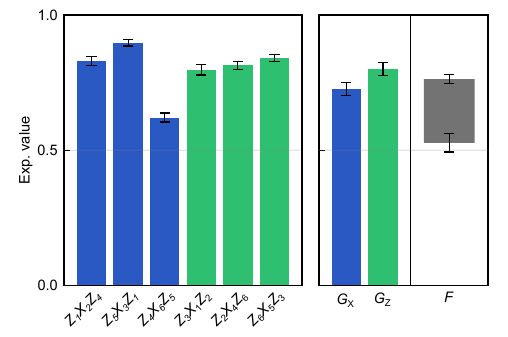}
    \caption{Fidelity estimation. The left plot contains the expectation values of the stabilizers $K_i$ with $i=1,\dots,6$. Each of these stabilizers acts on a different combination of three qubits, shown on the $x$-axis (e.g, $K_1=Z_3X_1Z_2$). Blue (green) bars correspond to stabilizers that can be measured with the logical $X$($Z$) measurement pattern. The right plot shows the estimation of the fidelity (gray box), alongside the expectation values of the operators $G_X$ (blue bar) and $G_Z$ (green bar). All the error bars represent a $1\sigma$ error.}
    \label{figure:fidelity_estimation}
\end{figure}

\subsubsection*{Entanglement witness for the hexagonal graph state}

We estimate the fidelity $F$ of the hexagonal graph state using a lower and an upper bound ($F_-$, $F_+$), both defined in terms of the stabilizers $K_{i}=X_{i}\prod_{k\in\mathcal{N}_{i}}Z_{k}$ (defined in the main text). The results are shown in \Fig \ref{figure:fidelity_estimation}. Since we measure the graph state with two measurement patterns, one for the logical $X$ and one for the logical $Z$ measurement, we define the two operators $G_X=\prod_{i=2,3,6}\frac{1+K_i}{2}$ and $G_Z = \prod_{i=1,4,5}\frac{1+K_i}{2}$, each one measurable only in the logical $X$ or logical $Z$ basis, respectively. The fidelity lower and upper bounds are defined~\cite{Thomas2024} by the equation

\begin{equation}
    \langle G_X \rangle + \langle G_Z \rangle -1 \leq F \leq \sqrt{\langle G_X \rangle\cdot\langle G_Z\rangle}.
\end{equation}

\noindent In this experiment, considering the events where we detect all photons, we measure a fidelity lower bound of $F_-=0.528(34)$ and an upper bound of $F_+=0.763(17)$.

\subsection*{Detailed graph code and loss-tolerance discussion}

\subsubsection*{Graph code general description}

In this section, we provide more details on the general construction of a graph code, which closely follows that of Refs.~\cite{Bell2023, Pettersson2025}. The stabilizer generators and logical operators of an $n-1$ qubit graph code can be derived from the \textit{progenitor graph}. The progenitor graph is an $n$ qubit graph state with a dedicated encoding qubit $E$, which, when measured, encodes the graph code composed of the remaining $n-1$ qubits. The logical $X$ operator is defined from the structure of the progenitor graph as $\overline{X} = \prod_{l \in \mathcal{N}_E}Z_l$, and the logical $Z$ is defined by $\overline{Z}=K_J\setminus Z_E$ for any qubit $J$ in the neighborhood of the encoding qubit $\mathcal{N}_E$. Furthermore, the stabilizer generators of the graph code are derived from the stabilizer generators of the progenitor graph as $\mathcal{K} = \{ K_JK_j, K_i \}_{j \in \mathcal{N}_E \setminus J, i \notin \mathcal{N}_E}$.

The above code construction can be explained differently by noting that the progenitor graph is a Bell state between the encoding qubit and the graph code $\ket{\Psi} = (\ket{0}_E\ket{\overline{0}} + \ket{1}_E\ket{\overline{1}})/\sqrt{2}$. Now, the logical operators can be seen from the equivalence of the progenitor graph stabilizer generators and the Bell-state stabilizer generators. That is, logical $X$ is derived from the stabilizer of the progenitor graph with $X$ support on the encoding qubit, and logical $Z$ is derived from a stabilizer with $Z$ support on the encoding qubit. More explicitly, the equivalence of the stabilizer $K_E = X_E\prod_{j \in \mathcal{N}_E}Z_j$ of the progenitor graph and the stabilizer $XX$ of the Bell state thus gives logical $X$, and the equivalence of the stabilizer $K_{j\in \mathcal{N}_E}$ and the stabilizer $ZZ$ of the Bell state gives logical $Z$.

\begin{figure}
    \centering
    \includegraphics[]{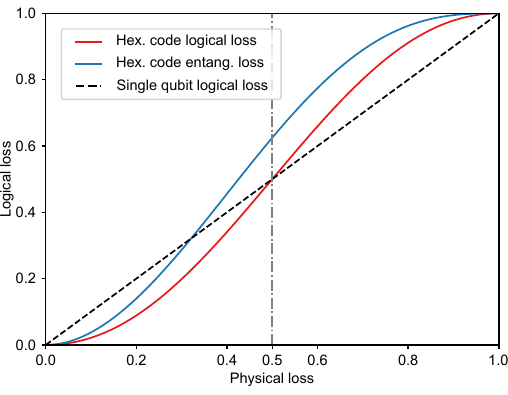}
    \caption{Logical loss theory curves. The black dashed line indicates the logical information loss for a single physical qubit. The red and blue curves show two theoretical models describing the logical loss in the hexagonal graph state. In the red curve, all successful loss configurations are included. In the blue curve, only configurations that are successful in both $X$ and $Z$ are considered, making it harder to beat the single qubit transmission (break-even point at $32\%$ physical loss instead of $50\%$).}
    \label{figure:log_loss_theo}
\end{figure}

\subsubsection*{Entanglement rate}
In our experiment, we perform loss-tolerant logical Pauli measurements of the Pauli $X$ and Pauli $Z$ operators. Both of these logical Pauli measurements have a logical loss of (red curve in \Fig \ref{figure:log_loss_theo}):

\begin{equation}
    \label{eq:log_trans}
    \overline{\ell} = 1 - (1-\ell)^5 - 5(1-\ell)^4\ell - 8 (1-\ell)^3\ell^2 - 2(1-\ell)^2\ell^3.
\end{equation}

\noindent This allows us to extract the logical $X$ or logical $Z$ information more efficiently than a single photonic qubit when the losses are below $50\%$. However, this efficiency advantage alone does not guarantee a higher rate of correct outcomes, since photon efficiency can be traded against fidelity. For example, we estimate that reducing the average physical photon loss from the current $\sim48\%$ to $45\%$ would require a fidelity of approximately $93\%$ to obtain a higher correct-outcome rate than for a single physical qubit. Furthermore, there are certain loss patterns for which we cannot claim entanglement between the spin qubit and the logical qubit. These are loss patterns where only logical $X$ or $Z$ can be retrieved. Labeling the photonic qubits from $1$ to $5$ as in the main text, the combinations for which we can retrieve only logical $X$ are $\{(1,2,3),(1,4,5),(2,3),(4,5)\}$, while for logical $Z$ they are $\{(1,2,4),(1,3,5),(2,4),(3,5)\}$. This leads us to define two different rates, the logical rate defined in Eq.~\eqref{eq:log_trans}, and an entanglement rate where we can measure both $X$ and $Z$ (blue curve in \Fig \ref{figure:log_loss_theo}), given by 

\begin{equation}
    \overline{\ell}_{XZ} = 1 - (1-\ell)^5 - 5(1-\ell)^4\ell - 6(1-\ell)^3\ell^2.
\end{equation}

\noindent In other words, in Eq.~\eqref{eq:log_trans} we do not solely rely on entanglement to retrieve the logical operators, but also on classical correlations. This might seem strange at first, but it is often used in error correction, where, for example, in fusion-based quantum computation, information from fusion failure (i.e., classical information from single-qubit Pauli measurements) greatly helps push the thresholds~\cite{Bartolucci2023}. Another example where this classical information is important is for one-way repeaters, where logical Pauli measurements are a key ingredient~\cite{Azuma2015, Pettersson2025}. Below, we furthermore give a detailed discussion of how the inclusion of classical correlations improves the threshold for loophole-free violation of Bell's inequality (not as a replacement for entanglement, but compared to complete erasure).

\subsubsection*{Logical qubit derivation}
To verify entanglement between the atomic qubit and the encoded qubit, we use an entanglement criterion based on qubits. Here we show that the logical operators $\overline{X}$ and $\overline{Z}$, for the measurement patterns where we can measure both (configurations in the left panel of Fig. 4 in the main text), obey the Pauli algebra on that subspace. This implies that, for those measurement configurations, the encoding system can be considered equivalent to a two-level logical qubit. In order to see this, we describe the logical operators in projector space for the different valid measurement patterns. For logical $X$ we obtain 
\begin{widetext}
\begin{align}
\label{eq:Xexp}
    \overline{X} = &\prod_{1, 2, 3, 4, 5}\mathbb{1}_1\mathbb{1}_2\mathbb{1}_3Z_4Z_5 + \prod_{1, 2, 3, 4} \mathbb{1}_1X_2X_3\mathbb{1}_4 + \prod_{1, 2, 3, 5} \mathbb{1}_1X_2X_3\mathbb{1}_5 \\
      + & \prod_{1, 2, 4, 5} \mathbb{1}_1\mathbb{1}_2Z_4Z_5 + \prod_{1, 3, 4, 5} \mathbb{1}_1\mathbb{1}_3Z_4Z_5 + \prod_{2, 3, 4, 5} \mathbb{1}_2\mathbb{1}_3Z_4Z_5 \notag \\
     + & \prod_{2, 3, 4} X_2X_3\mathbb{1}_4 + \prod_{1, 3, 4} Z_1X_3Z_4 + \prod_{2, 3, 5} X_2X_3\mathbb{1}_5 + \prod_{1, 2, 5} Z_1X_2Z_5 \notag \\
     + & \prod_{2, 4, 5}\mathbb{1}_2Z_4Z_5 + \prod_{3, 4, 5}\mathbb{1}_3Z_4Z_5 \notag
\end{align}
\end{widetext}
\noindent and for logical $Z$
\begin{widetext}
\begin{align}
\label{eq:Zexp}
    \overline{Z} = & \prod_{1, 2, 3, 4, 5}\mathbb{1}_1\mathbb{1}_2Z_3\mathbb{1}_4X_5 + \prod_{1, 2, 3, 4} \mathbb{1}_1Z_2\mathbb{1}_3X_4 + \prod_{1, 2, 3, 5} \mathbb{1}_1\mathbb{1}_2Z_3X_5 \\
    + & \prod_{1, 2, 4, 5} X_1Z_2\mathbb{1}_4X_5 + \prod_{1, 3, 4, 5} \mathbb{1}_1Z_3\mathbb{1}_4X_5 + \prod_{2, 3, 4, 5} \mathbb{1}_2Z_3\mathbb{1}_4X_5 \notag \\
    + & \prod_{2, 3, 4} Z_2\mathbb{1}_3X_4 + \prod_{1, 3, 4} X_1Z_3X_4 + \prod_{2, 3, 5} \mathbb{1}_2Z_3X_5 + \prod_{1, 2, 5} X_1Z_2X_5 \notag \\
    + & \prod_{2, 4, 5}Z_2X_4\mathbb{1}_5 + \prod_{3, 4, 5}Z_3\mathbb{1}_4X_5 \notag,
\end{align}
\end{widetext}
\noindent and for logical $Y$
\begin{widetext}
\begin{align}
    \overline{Y} = i\overline{X}\overline{Z} = & \prod_{1, 2, 3, 4, 5}\mathbb{1}_1\mathbb{1}_2Z_3Z_4Y_5 - \prod_{1, 2, 3, 4} \mathbb{1}_1Y_2X_3X_4 - \prod_{1, 2, 3, 5} \mathbb{1}_1X_2Y_3X_5 \\
    + & \prod_{1, 2, 4, 5} X_1Z_2Z_4Y_5 + \prod_{1, 3, 4, 5} \mathbb{1}_1Z_3Z_4Y_5 + \prod_{2, 3, 4, 5} \mathbb{1}_2Z_3Z_4Y_5 \notag \\
    - & \prod_{2, 3, 4} Y_2X_3X_4 + \prod_{1, 3, 4} Y_1Y_3Y_4 - \prod_{2, 3, 5} X_2Y_3X_5 + \prod_{1, 2, 5} Y_1Y_2Y_5 \notag \\
    + & \prod_{2, 4, 5}Z_2Y_4Z_5 + \prod_{3, 4, 5}Z_3Z_4Y_5 \notag.
\end{align}
\end{widetext}
\noindent Here, $\prod_{\{i\}}$ is the projector onto the sub-space of the non-lost photons configuration $\{i\}$, and it is assumed that the sub-space projector projects with vacuum $\ket{\varnothing}\bra{\varnothing}$ on the lost photons. From the above expressions, it can be readily obtained that $\overline{X}^2 = \overline{Z}^2 = \overline{Y}^2 = \sum_{E}\prod_E\overline{\mathbb{1}}$, where the sum goes over all projectors ($\prod_E$) onto the subspace where we can measure both logical $X$ and $Z$. Furthermore, we see that the logical operators indeed anticommute $\{ \overline{X}, \overline{Z} \} = 0$, and that they obey the commutation relation $[\overline{Z}, \overline{X} ] = 2i\overline{Y}$. This can be seen by looking at all sub-space projectors separately, as the multiplication of different sub-space projectors yields $0$. For instance, for the $\prod_{1, 2, 3, 4, 5}$ projector, the anticommutator $\{ \overline{X}, \overline{Z} \}$ gives
\begin{widetext}
\begin{align}\mathbb{1}_1\mathbb{1}_2\mathbb{1}_3Z_4Z_5 \otimes \mathbb{1}_1\mathbb{1}_2Z_3\mathbb{1}_4X_5 + \mathbb{1}_1\mathbb{1}_2Z_3\mathbb{1}_4X_5 \otimes \mathbb{1}_1\mathbb{1}_2\mathbb{1}_3Z_4Z_5 \notag \\
    = -i\mathbb{1}_1\mathbb{1}_2Z_3Z_4Y_5 + i\mathbb{1}_1\mathbb{1}_2Z_3Z_4Y_5 = 0,
\end{align}
\end{widetext}
\noindent and similarly for the commutator $[\overline{Z}, \overline{X} ]$ we get
\begin{widetext}
\begin{align}
    \mathbb{1}_1\mathbb{1}_2Z_3\mathbb{1}_4X_5 \otimes \mathbb{1}_1\mathbb{1}_2\mathbb{1}_3Z_4Z_5 - \mathbb{1}_1\mathbb{1}_2\mathbb{1}_3Z_4Z_5 \otimes \mathbb{1}_1\mathbb{1}_2Z_3\mathbb{1}_4X_5 \notag \\
    =i\mathbb{1}_1\mathbb{1}_2Z_3Z_4Y_5 + i\mathbb{1}_1\mathbb{1}_2Z_3Z_4Y_5  =2i \times (\mathbb{1}_1\mathbb{1}_2Z_3Z_4Y_5), 
\end{align}
\end{widetext}
\noindent which is $2i$ times the $\prod_{1, 2, 3, 4, 5}$ sub-space projector of logical $Y$. These results verify that the logical operators obey the Pauli-algebra on the subspace where we can detect both logical $X$ and $Z$ and thus define a qubit degree of freedom on this subspace.

The expressions above considered only the measurement configurations where both $\overline{X}$ and $\overline{Z}$ are recoverable. If we add the configurations where only one between $\overline{X}$ or $\overline{Z}$ is recoverable (center and left panel of Fig. 4 in the main text, respectively) to their respective expansions in Eqs.~\eqref{eq:Xexp} and \eqref{eq:Zexp}, the commutation relations are, however, no longer valid. In particular, these added terms do not anticommute. In other words, the graph code no longer defines a qubit, as it only has access to a single logical operator. We note that this logical operator still contains (classical) correlations on a specific basis, which can have practical use in potential applications (see the next section). On the other hand, in the context of state validation using the criterion from Tóth and Gühne~\cite{GuhneToth2005}, where two-level systems are assumed, this issue implies that we need to invoke a \textit{fair sampling assumption} for these additional configurations in order to match the considered scenario. Namely, we have to assume that the statistics obtained in the additional configurations are still representative of the statistics obtained in the other configurations compatible with the two-level system description. In other words, we assume that there is no adversarial agent that counterfeits the system for these particular photon loss patterns.

\subsubsection*{CHSH inequality with a logical qubit}

A concrete application where the classical correlations discussed above would prove useful, is in a loophole free Clauser–Horne–Shimony–Holt (CHSH)~\cite{CHSH} inequality experiment. Given the entangled state $\ket{\psi} = \alpha \ket{00} + \beta \ket{11}$ shared between Alice and Bob, the CHSH inequality is given by

\begin{equation}
    \label{eq:CHSH}
    S = |\langle ab \rangle + \langle ab' \rangle + \langle a'b \rangle - \langle a'b' \rangle| \leq 2,
\end{equation}

\noindent where $a$ ($a'$) and $b$ ($b'$) represent different measurement bases for Alice and Bob. Taking into account the loss of photonic qubits, we can, as in Ref.~\cite{Gonzalez24}, define $a = 2P_+ - 1$, where $P_+$ is the plus projection operator of the given basis $a$, i.e., $P_+ = \eta\frac{1 + \Vec{n} \cdot \Vec{\sigma}}{2} = \eta\frac{1 + \cos{\theta_a} \sigma_x + \sin{\theta_a}\sigma_z}{2}$, with $\eta$ being the transmission efficiency of the qubit. With this expression for $a$ (and $b$), we can find the expectation value $\langle ab \rangle$ as
\begin{widetext}
\begin{align*}
    \langle ab \rangle = \eta_A \eta_B (2\alpha \beta \cos{\theta_a}\cos{\theta_b} + \sin{\theta_a}\sin{\theta_b} + (\sin{\theta_a} + \sin{\theta_b}) (\alpha^2 - \beta^2) +1) \\
    -\eta_A(\sin{\theta_a}(\alpha^2 - \beta^2) + 1) - \eta_B(\sin{\theta_b} (\alpha^2 - \beta^2) + 1) + 1.
\end{align*}
\end{widetext}
In our experiment, Alice's qubit is an atom, and we thus for simplicity consider $\eta_A = 1$, corresponding to a deterministic measurement on the atom. $\eta_B$ corresponds to the logical efficiency of the photonic qubit. Furthermore, if Bob has the hexagonal logical qubit, he can only measure it in the logical $X$ and $Z$ basis by simple single-qubit measurements, and we thus take $\theta_b = \pi / 2$ and $\theta_{b'} = 0$.

\begin{figure}
    \centering
    \includegraphics[]{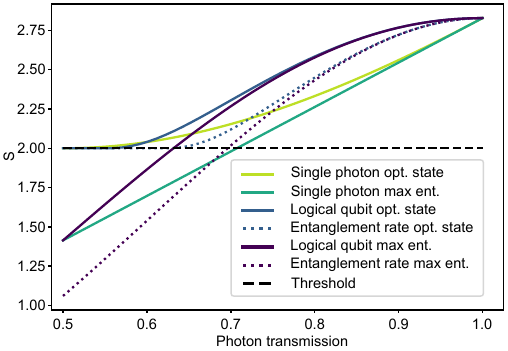}
    \caption{The various curves represent the $S$ value as a function of the photon transmission ($\eta$) for different states. We plot $S$ for Bob having a hexagonal logical qubit and a single photonic qubit. For each of these two cases, we plot $S$ for a maximally entangled state (max ent.) with $\alpha = \beta = 1/\sqrt{2}$ and for an optimized state (opt. state) over the parameters $\alpha$, $\beta$, and the measurement bases. The dashed line is the loophole-free Bell test threshold. Apart from a small region, where the violation is anyway small, the encoded qubit leads to a stronger violation of the CHSH inequality. For reference, we also plot the $S$ value when using the entanglement rate for Bob's logical qubit, i.e., where we only accept events where we can measure both logical $X$ and logical $Z$. This leads to a lower value of $S$. The inclusion of the classically correlated events thus improves the violation of the CHSH inequality.}
    \label{fig:bell_curves}
\end{figure}

In \Fig \ref{fig:bell_curves}, we plot the $S$ value for a maximally entangled state $(\ket{00} + \ket{11}) / \sqrt{2}$, where Bob has either a single photon or a logical photonic qubit. Here, we see a clear advantage in using a logical qubit rather than a single photon, as it yields a higher value of $S$. Since $(ab)^2=1$ (and similarly for the other measured quantities), the variance is determined by the mean values. Therefore, a higher value of $S$, also in general, means a higher statistical significance. In \Fig \ref{fig:bell_curves}, we furthermore plot the $S$ value over photon transmission for Bob receiving a single photon or a logical photonic qubit, where at each point we optimize the state $\ket{\psi} = \alpha \ket{00} + \beta \ket{11}$ and Alice's measurement bases. For the single photon case, an optimization of Bob's measurement bases is also done, which is why the single photon works better than the logical qubit below $58 \%$ transmission. In this regime, however, the value of $S$ is limited, making it hard to verify the violation of the CHSH inequality. In practice, the creation of the logical state $\ket{\psi} = \alpha \ket{00} + \beta \ket{11}$ for our setup can be realized by measuring the first atom in the conjugate basis $\ket{\psi^*} = \beta^*\ket{0} + \alpha^*\ket{1}$. This can be seen by doing the projection $\ket{\psi^*}\bra{\psi^*}_{A_1}$ on the composite atom-atom and logical qubit state $\ket{\Psi} = (\ket{1}_{A_1}\ket{0}_{A_2}\ket{\overline{0}} + \ket{0}_{A_1}\ket{1}_{A_2}\ket{\overline{1}}) / \sqrt{2}$. The curves thus correspond to concrete strategies that can be implemented in our experimental setup, although the fidelities are currently not sufficient to observe any violation. 

Both for maximally and non-maximally entangled states, the logical rate, where we include configurations where we can only measure logical $X$ or logical $Z$, thus leads to a stronger violation of the CHSH inequality (apart from a tiny region, where the violation is anyway small). Had we only included the configurations, where we can measure both logical $X$ and logical $Z$ (corresponding to the entanglement rate), the difference between encoded and non-encoded would have been smaller, as can be seen in \Fig \ref{fig:bell_curves}.

\subsection*{Data acquisition and analysis}

\subsubsection*{Data acquisition}

The primary challenge during data acquisition was the drift of efficiency over time. On a time scale of a few hours ($\sim 3 - 5 \mathrm{h}$), the source-to-detection efficiency of the single photons generated via the addressing vSTIRAP beam decreased by $1-2\%$ from its nominal value. To prevent the efficiency from falling below the break-even point of $50\%$, we decided to interrupt the experiment approximately every $4 \mathrm{h}$ to measure the photon efficiency. Whenever the measured efficiency was significantly lower than expected, we retuned the system before resuming data collection for another $4\mathrm{h}$. In total, we acquired $\sim26\mathrm{h}$ of data in the $X$ basis and $\sim18\mathrm{h}$ in the $Z$ basis.

The duration of the experimental sequence, without considering the power stabilization and the laser cooling sequences, is $1.346$ms. The full experimental sequence duration is $3$ms.

\begin{table}
    \centering
    \begin{tabular}{|l|r|r|r|r|r|}
    \hline
      & Tot. & BSM filtered & Heralded & Successful & Full ring \\
    \hline
    $X$ & 39994 & 24590 & 8038 & 4356 & 377 \\
    $Z$ & 30620 & 19179 & 8000 & 4290 & 338 \\
    \hline
    \end{tabular}
    \caption{Number of events at various stages of the data analysis for the $X$ and $Z$ bases. In the first column we give the number of total events (already including the success of the first fusion, which is hard coded in the experimental sequence, and the second fusion). In the second column the events are filtered considering the time parameters of the two fusions. In the third column we only keep the events accordingly to the heralding criteria. In the fourth column we provide the number of events that characterize the logical loss, i.e, the number of events for which the loss configuration allows the reconstruction of the logical qubit. Finally, the fifth column contains the number of events in which all photons are detected.}
    \label{tab:number_events}
\end{table}

\subsubsection*{Data analysis and heralding criteria}

From the raw data, we filter out some events depending on the heralding criteria. We keep only the events where both fusions are successful, meaning the events where the two photons are detected at opposite detectors. We restrict the time difference between the two photons to values smaller than $250$ns and $400$ns for the first and second fusion, respectively. We always herald on the successful detection in the $X$ basis of the first photon (qubit 0). For the logical efficiency results, in the $X$ basis, the atomic qubits 6 and 7 must be both detected, while for the $Z$ basis, only one qubit (qubit 6) is necessary. In table \ref{tab:number_events} we give the number of total, filtered, and successful events for the two bases. The error bars are calculated using the normal approximation of the binomial proportion error bar.